%% file: main.tex
\documentclass[conference]{IEEEtran}
\IEEEoverridecommandlockouts

\newcommand{\ignore}[1]{}
\usepackage{microtype}
\usepackage[pass]{geometry}
\usepackage{fancyhdr}
\usepackage[normalem]{ulem}
\usepackage[hyphens]{url}
\usepackage{hyperref}
\usepackage{color}
\usepackage{soul}

\usepackage{mdframed}
\usepackage{lipsum}
\newmdtheoremenv{theo}{Definition}

\usepackage{subfig}
\usepackage{graphicx,wrapfig,lipsum}

\usepackage{tikz}

\newcommand*\circled[1]{\tikz[baseline=(char.base)]{
            \node[shape=circle,draw,inner sep=1pt] (char) {#1};}}

\newtheorem{obs}{Observation}

\renewcommand{\baselinestretch}{0.972}

\fancypagestyle{firstpage}{
  \fancyhf{}
\setlength{\headheight}{50pt}

  \fancyhead[C]{\normalsize{Submitted to arXiv}}
  \pagenumbering{arabic}
}

\title{Triad-NVM: Persistent-Security for Integrity-Protected and Encrypted Non-Volatile Memories (NVMs) \thanks{DISTRIBUTION A.   Approved for public release. Distribution unlimited. Case
Number 88ABW-2018-4218 Dated 29 Aug 2018.}}

\author{\IEEEauthorblockN{\textsuperscript{} Amro Awad}
\IEEEauthorblockA{
\textit{University of Central Florida}\\
Orlando, FL - USA \\
amro.awad@ucf.edu}
\and
\IEEEauthorblockN{\textsuperscript{} Laurent Njilla}
\IEEEauthorblockA{
\textit{Air Force Research Laboratory}\\
Rome, NY - USA \\
laurent.njilla@us.af.mil}
\and
\IEEEauthorblockN{\textsuperscript{} Mao Ye}
\IEEEauthorblockA{
\textit{University of Central Florida}\\
Orlando, FL - USA \\
mye@knights.ucf.edu}
}

\clearpage

\begin{document}
%\maketitle

%\begin{center}

%\title{Triad-NVM: Persistent-Security for Integrity-Protected and Encrypted Non-Volatile Memories (NVMs)}

\clearpage
\twocolumn

%\author{%
%  Amro Awad {(University of Central Florida, Orlando FL)}%
%  \and Laurent Njilla {(Air Force Research Laboratory, Rome NY)}%
%  }

\maketitle
\thispagestyle{firstpage}
\pagestyle{plain}

%%%%%% -- PAPER CONTENT STARTS-- %%%%%%%%

\input{abstr}

\input{intro}
\input{background}
\input{design}

\input{metho}
\input{evaluation}

\input{relat}
\input{concl}
%%%%%%% -- PAPER CONTENT ENDS -- %%%%%%%%

%%%%%%%%% -- BIB STYLE AND FILE -- %%%%%%%%
\bibliographystyle{ieeetr}
\bibliography{ref}
%%%%%%%%%%%%%%%%%%%%%%%%%%%%%%%%%%%%

\end{document}

%% file: abstr.tex
\begin{abstract}

Emerging Non-Volatile Memories (NVMs) are promising contenders for building future memory systems. On the other side, unlike DRAM systems, NVMs can retain data even after power loss and thus enlarge the attack surface. While data encryption and integrity verification have been proposed earlier for DRAM systems, protecting and recovering secure memories becomes more challenging with persistent memory. Specifically, security metadata, e.g., encryption counters and Merkle Tree data, should be securely persisted and recovered across system reboots and during recovery from crashes. Not persisting updates to security metadata can lead to data inconsistency, in addition to serious security vulnerabilities.

In this paper, we pioneer a new direction that explores persistency of both Merkle Tree and encryption counters to enable secure recovery of data-verifiable and encrypted memory systems. To this end, we coin a new concept that we call \textit{Persistent-Security}. We discuss the requirements for such persistently secure systems, propose novel optimizations,  and evaluate the impact of the proposed relaxation schemes and optimizations on performance, resilience and recovery time. To the best of our knowledge, our paper is the first to discuss the persistence of security metadata in integrity-protected NVM systems and provide corresponding optimizations. We define a set of relaxation schemes that bring trade-offs between performance and recovery time for large capacity NVM systems. Our results show that our proposed design, Triad-NVM, can improve the throughput by an average of ~2x (relative to strict persistence). Moreover, Triad-NVM  maintains a recovery time of less than 4 seconds for an 8TB NVM system (30.6 seconds for 64TB), which is ~3648x faster than a system without security metadata persistence.

\end{abstract}

%% file: intro.tex
\section{Introduction}

Securing Non-Volatile Memory (NVM) systems is a critical requirement for their deployment. Unlike DRAM, NVMs retain data after powering off the system, which necessiates encrypting them to avoid data remenance attacks. Moreover, NVMs enable persistent applications which can recover after a crash by frequently persisting their data on NVMs. However, while both concepts, persistency and security, seem orthogonal, they become naturally relevant with emerging NVMs; the system should be able to recover securely and guarantee its data integrity and confidentiality across crashes. Unfortunately, such synergy between persistency and security has received limited attention from the research community, which we believe is due to the following reasons. First, securing memory systems, e.g., encryption and data integrity verification, has been originally developed for DRAM systems to defeat cold-boot attacks \cite{CB1,CB2,CB3}. In such systems, persistency and data recovery is not anticipated and thus all security metadata are expected to be reinitialized \cite{ISCA_DRAM}. Second, the research on encrypting and securing NVMs has been mostly focused on performance of normal-operation without consideration of maintaining consistency across crashes \cite{SYNERGY,SS,DEUCE,ASSURE}. Third, persisting data in emerging NVMs has many compelling models, notably JUSTDO logging \cite{JUSTDO} and epoch-based persistence \cite{THYNVM},   where the focus is on data persistency than its accompanying security metadata \cite{LOG1,LOG2,LOG3,LOG4}. 

%While prior research work has assumed that systems have sufficient residual or backup power to flush encryption metadata, i.e., encryption counters, in the event of power loss \cite{SS,OBFUSMEM}. 

%Although feasible in theory, in practice, reasonably long-lasting Uninterruptible Power-Supplies (UPS) are expensive and occupy large areas. Admittedly, 

While the Asynchronous DRAM Refresh (ADR) feature has been a requirement for many NVM form factors, e.g., NVDIMM \cite{chang2015nvdimm}, most processor vendors limit persistent domains within the processor chip to tens of entries in the \textit{write pending queue (WPQ)}\cite{rudoff2016dpi}. The main reason is the high costs for ensuring a long-lasting sustainable power supply in case of power loss in addition to the power-consuming nature of NVM writes. As security metadata can be hundreds of kilobytes or megabytes, the system-level ADR support would fail to guarantee the persistence of all their updated values within the processor chip. For many users, affording powerfull backup batteries or deploying expensive power-supplies is infeasible, either due to environmental, limited area or cost limitations, thus battery-free solutions are always in demand\cite{OSIRIS}.

Recently, several work have explored persisting encryption counters on secure NVMs \cite{HPCA_KHAN,OSIRIS,SECPM}. Unfortunately, none of the previous work investigate persisting integrity protected systems and its accompanying metadata, but rather limit their schemes to encryption counters only. In fact, persisting integrity-verification metadata has much higher overhead. Moreover, most of the previous work do not clearly define which security metadata need to be persisted and the impact of relaxing the persistence of such metadata on security, recovery time, resilience, performance and write endurance. In summary, the current literature in persisting security metadata lacks the following: \circled{1} Solid definition of what data needs to be persisted in secure memory systems, especially for systems with integrity protection. \circled{2} Complete understanding of the impact of relaxing security persistence, e.g., selective counter persistence \cite{HPCA_KHAN}, on security and best practices that \textit{must} accompany such relaxation schemes. \circled{3} Understanding of the impact of different relaxation schemes on recovery time and resilience of the system, i.e., possibility of recovery failure. To address all these shortcomings, we first discuss the problem of persisting security metadata in integrity-protected secure systems. Later, we define a new concept, \textit{\textbf{Persistent Security}}, which defines the requirements for \textit{securely} recovering integrity-protected and encrypted memory systems. Finally, we discuss several relaxation schemes and a comprehensive design that leverages them to optimize performance.

%As emerging NVMs are expected to provide capacities in order of terabytes, the number of levels of the Merkle Tree, commonly used to protect data integrity, is expected to be in orders of 10-20 levels. Moreover, NVMs have limited write endurance which gets exacerbated due to encryption avalanche effect \cite{DEUCE,SS}. However, to ensure consistent view of memory after recovery, such metadata need to be persisted and recovered after crashes or system reboots. However, since each data block being written back to memory will result in an encryption counter update and updating all levels of Merkle Tree (10-20 updates), persisting all these updates on each write operation can significantly degrade the performance of secure NVM systems. 

 To reduce the overhead of persisting Merkle Tree and encryption counters after each update, we propose \textit{Triad-NVM}, a novel design that leverages four key insights: \circled{1} Current systems and support for NVM clearly define regions of the NVM device that can be used as persistent memory, e.g., initializing the Linux kernel with the parameter {\tt memmap=4G!12G}, can be used to dedicate the 4GB starting from address 12GB to be mounted/mapped as persistent memory. \circled{2} Being able to spatially divide the address space into persistent and non-persistent regions allows to partition Merkle Tree vertically into persistent and non-persistent subtrees. While prior work \cite{HPCA_KHAN} distinguish between persistent and non-persistent data through modifying applications to explicitly hint hardware and memory controller, we observe that such modifications might be not needed if persistent regions are well-defined and spatially distinguishable. \circled{3} Merkle Tree can be reconstructed after a crash by only ensuring the persistence of low levels, however, at the cost of resilience (single-point of failure) and recovery time. \circled{4} Building Merkle Tree for Non-Persistent regions at recovery time can be very expensive, however, it can be mitigated by encoding special values at some levels of the Merkle Tree during recovery.

To evaluate our design, we use Gem5 \cite{GEM5}, a full-system cycle-accurate simulator. We use Linux kernel version 4.14 along with a disk image based on Ubuntu 16.04. Additionally, we initialize the kernel to dedicate the last 4GB (out of 16GB) as a persistent region, which we later mount as a directly-accessible (DAX) ext4 filesystem. The persistent region can be used by any library that supports persistent memory, however, we opt for using Intel's PMDK library \cite{PMDK} (previously known as PMEM) to build 3 microbenchmarks, in addition to 4 DAX-based synthetic workloads, along with 12 benchmarks from the SPEC2006 suite \cite{SPEC}. We additionally evaluate 4 workloads that run combinations of persistent (PMDK and DAX)  and non-persistent (SPEC) workloads. Our simulation results show that Triad-NVM can improve the throughput by an average of ~2x (relative to strict persistence). Moreover, Triad-NVM  maintains a recovery time of less than 4 seconds for an 8TB NVM system (30.6 seconds for 64TB), which is ~3648x faster than a system without security metadata persistence.

\begin{figure*}[htbp!]
\begin{center}
  \vspace{-1em}
   \includegraphics[scale=0.45]{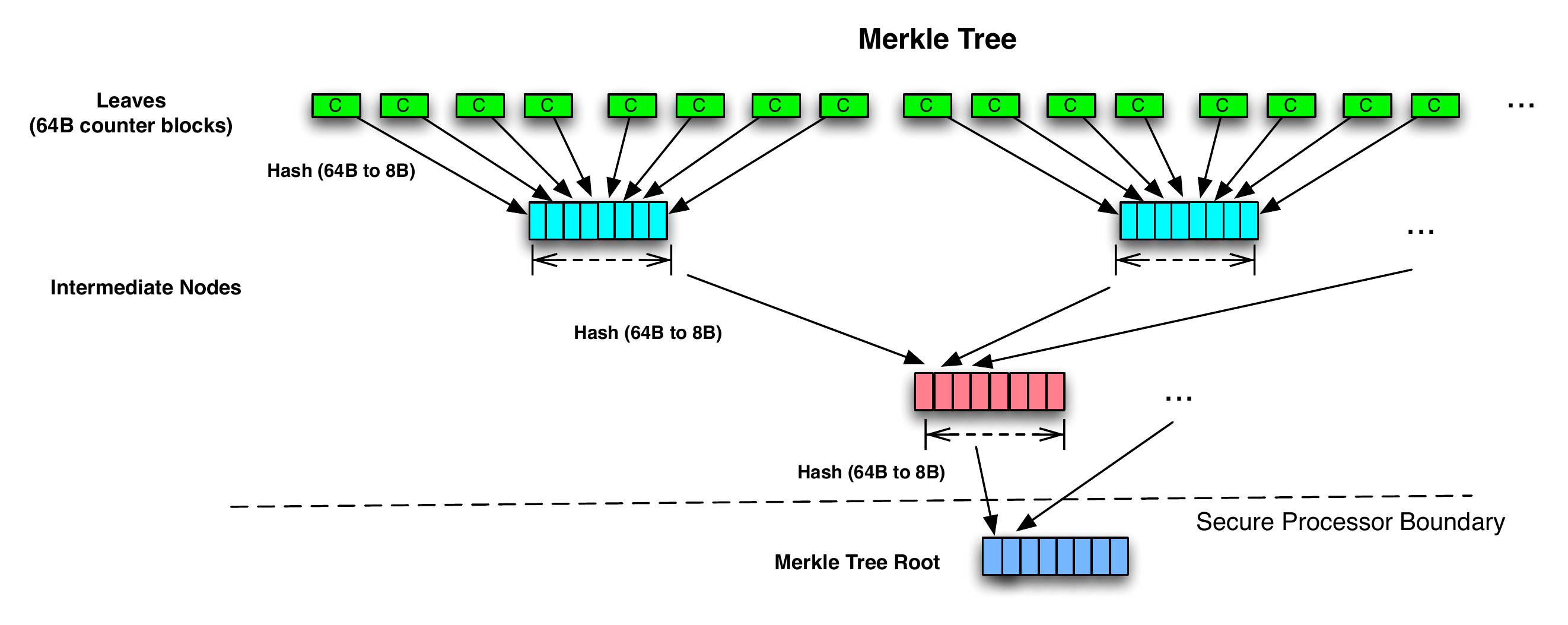}
  \caption{Example Merkle Tree.} %State of the art counter mode encryption. AES is shown for encryption/decryption but other cryptographic algorithms are possible.}
  \vspace{-1em}
  \centering
  \label{fig:MT}
  \end{center}
\end{figure*}

The rest of the paper is organized as follows. First, we discuss the problem and background of secure NVM systems in Section \ref{sec:back}, where we also discuss the conventional way of persisting security metadata in secure NVMs. Later, in Section \ref{sec:design}, we discuss the requirements of strictly-persistent secure memory systems. Later, we discuss different relaxation schemes and their impact on security, performance and resiliency of NVM systems. In Section \ref{sec:metho}, we discuss our evaluation methodology. Our results and discussion of our evaluation are presented in Section \ref{sec:eval}. Section \ref{sec:relat} discusses the most relevant work to our paper. Finally, we conclude our work in Section \ref{sec:concl}.

%% file: background.tex
\section{Background and Motivation}
\label{sec:back}

We will start our section with background discussion followed by demonstrating the issue of crash consistency of security metadata. Later, we present motivational data of the impact of the problem.

\subsection{Background}

Emerging Non-Volatile Memories (NVMs), such as Intel's and Micron's 3D XPoint \cite{3DXPOINT}, are just around the corner and they are expected to be on the market soon. Unlike DRAM, they feature high-density, ultra-low idle power (no need to refresh cells), low latency and persistence of data \cite{PCM1,PCM2,PCM3}. Due to their ability to retain data across system reboots, they can be also used to store files and persistent data structures. Hence, NVMs can be used as a memory and a storage device at the same time. Specifically, NVM-based DIMMs can be used to hold files and also regular  memory pages, and can be accessed in a way similar to DRAM through load/store operations. To realize this, new Operating Systems (OSes) started to support configuring the memory as both filesystem and conventional memory. In particular, recent linux kernels and Windows started to support directly-access (DAX) support filesystems \cite{DAX}. In DAX supported filesystems, e.g., ext4 with DAX, a file can be directly memory-mapped and accessed through typical load/store operations, however, without the need to copy its pages to page cache as in conventional filesystems. For example, the NVM memory can be configured to have a part dedicated to hold filesystem. In linux systems, when initializing the kernel, the parameter {\tt memmap=4G!12G} can be used to dedicate the 4GB starting from address 12GB to directly-accessible filesystems. Thus, the same memory chip will be accessed for both files and conventional memory pages.

\subsubsection{NVM Security} Due to non-volatility, emerging NVMs facilitate data remenance attacks; data remains there even after losing power. Accordingly, emerging NVMs are commonly coupled with memory encryption and data integrity verification. State-of-the-art secure NVM systems use counter-mode encryption\cite{SS,DEUCE,HPCA_KHAN}. Counter-mode encryption can protect against replay attacks, scanning memory and bus snooping attacks. Counter-mode memory encryption associates a counter with each 64B block in memory and this counter changes each time the corresponding block is being written to memory and the new counter value will be used to do the encryption. The most-recent value of the counter must be book-kept to be later used for decryption. In counter-mode encryption, repeating the same counter value can result in serious security vulnerabilities, e.g., known-plaintext attacks. To avoid replaying counters, they are typically protected against tampering through Merkle Tree. State-of-the-art work, e.g., Bonsai Merkle Tree \cite{ISCA_DRAM}, applies Merkle Tree over the encryption counters and protects data through computing MAC value over ciphertext and data. 

As shown in Figure \ref{fig:MT}, each encryption counter, which is associated with a data block, is used to calculate a hash value (MAC) that will be used along with other hash values from other groups of counters to create a lower level hash value. Finally, the resulting MAC value, which is called \textit{root}, is kept in the processor. Each time a counter is brought from the insecure area, e.g., memory module, it will be verified by calculating the upper hash values and see if the result matches the root kept in the processor. Furthermore, when the processor writes a memory block and updates the corresponding counter, the Merkle Tree intermediate nodes and root should be updated to reflect the most-recent change.

\subsubsection{Memory Counter-Mode Encryption}  
\begin{figure}[htp!]
\begin{center}
  \vspace{-0.4em}
   \includegraphics[scale=0.35]{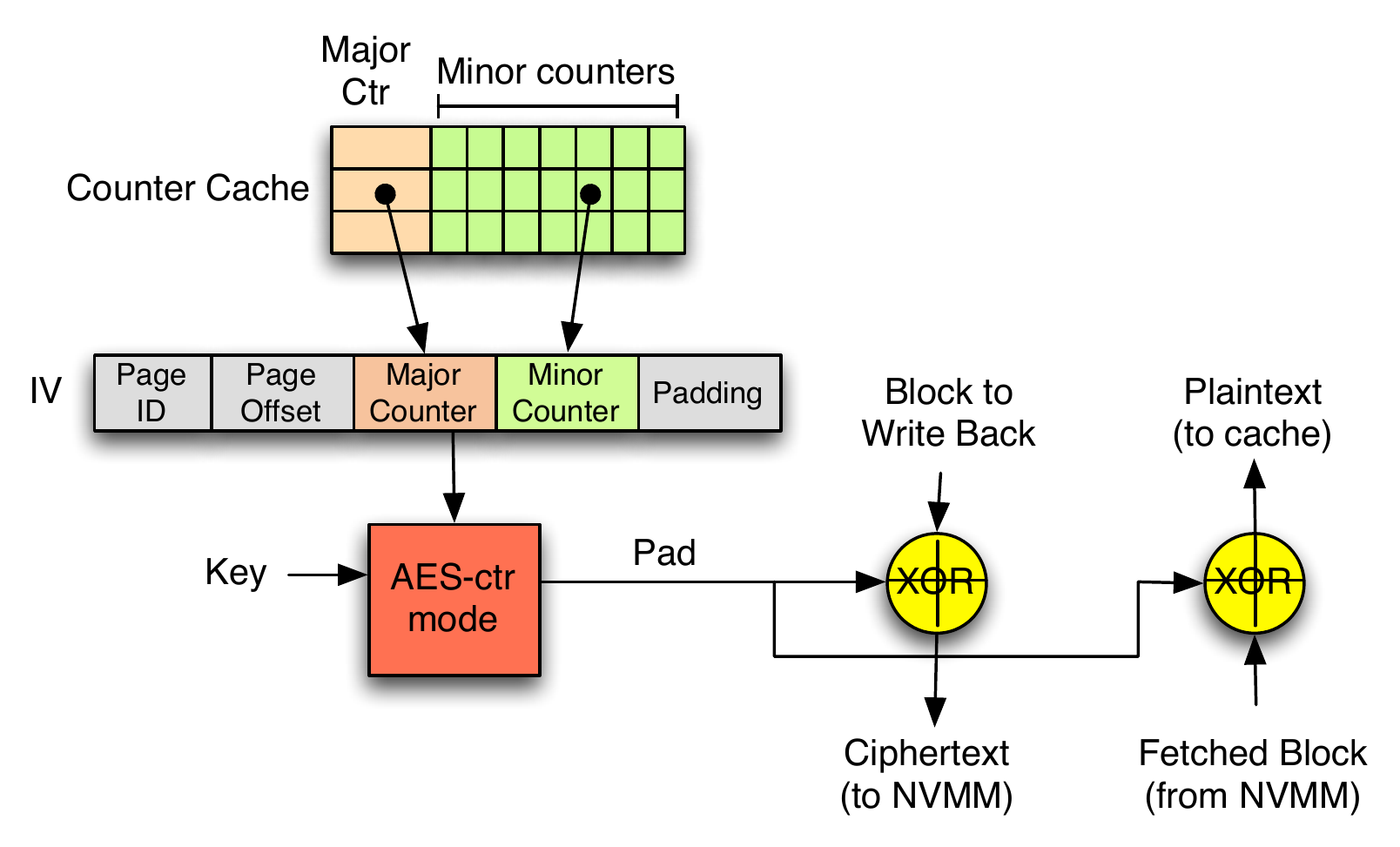}
  \caption{State-of-the-art counter mode encryption as in \cite{SS}}
  \centering
  \vspace{-1em}
  \label{fig:countermode}
  \end{center}
\end{figure}

In counter-mode encryption, the encryption algorithm, e.g., AES, takes {\em initialization vector} (IV) as its input to create a one-time pad (OTP) as explained in Figure~\ref{fig:countermode}. Later, when the block arrives the processor chip, a low-cost bitwise XOR with the pad (encrypted IV) is needed to obtain the plaintext. By doing so, the decryption latency is hidden by the memory access latency. In our paper, we use state-of-the-art design for organizing encryption counter, split-counter scheme  \cite{SS,ISCA_DRAM}, where each IV consists of a unique ID of a page (to distinguish between swap space and main memory space), page offset (to guarantee different blocks in a page will get different IVs), a per-block {\em minor} counter (to make the same value encrypted differently when written again to the same address), and a per-page {\em major} counter (to guarantee uniqueness of IV when minor counters overflow).

% Since the latter approach provides stronger security (the former approach only protects against data remanence attacks), we assume the latter approach.

As in previous work~\cite{SS,HPCA_KHAN,OSIRIS,MERKLEE, ISCA_DRAM, DEUCE}, we assume counter-mode memory encryption. In addition to its performance advantages, it also provides strong security defenses against a wide range of attacks. Counter-mode encryption is secure against dictionary-based attacks, known-plaintext attacks, bus snooping and replay attacks. In split-counter scheme, the encryption counters are organized as major counters (shared between cache blocks of the same page) and minor counters that are specific for each cache block \cite{ISCA_DRAM}. Such organization allows packing 64 cache blocks' counters in a 64B block; 7-bit minor counters and 64-bit major counter. Major counters are incremented when one of its minor counters overflows, in which all corresponding minor counters will be reset and the whole page will be re-encrypted using the new major counter\cite{ISCA_DRAM}. When the major counter of a page overflows (64-bit counter),  a new key is generated and the memory contents will be re-encrypted using the new key. Split-counter scheme provides significant reduction of memory re-encryption rate and minimizes the storage overhead of encryption counters when compared to other schemes, e.g., monolithic counter scheme or independent counter for each cache block. Moreover, split-counter exploits the spatial locality of encryption counters, and hence achieves a higher counter cache hit rate. Similar to state-of-the-art work \cite{SS,OBFUSMEM,SYNERGY,ISCA_DRAM}, we use split-counter scheme for organizing the encryption counters.

\subsubsection{Impact of Persistency on Security} As persistent memory can be utilized to store checkpoints and recover from crashes, the accompanying security metadata should be persisted too. For instance, we mentioned earlier that the counter used for encryption will be updated on each write operation of the associated memory block, however, such update can occur on a volatile \textit{counter cache} inside the processor chip, hence not persisted to the NVM memory. Accordingly, the data block could be encrypted with the new counter value and written to NVM, but the counter is not updated in memory. Clearly, if a crash occurs, a stale counter value will be used to do the decryption, which will result in incorrect decryption. While the recovery of non-persistent data is not important, it is \textit{critical} to avoid reusing their counters' previous values; repeating an old (not persisted) counter will result in a reuse of OTP that could have been observed earlier, which results in security vulnerabilities as described below. %Accordingly, it is necessary to also persist encryption counters \cite{HPCA_KHAN}. However, while prior work selectively persist counters only for persistent data structures \cite{HPCA_KHAN}, this can lead to security vulnerability. It is crucial to never repeat a counter a value even for non-persistent (not part of DAX) memory locations or data structures. Below is a discussion of a potential attack for not persisting counters of non-persistent data structures.

\subsubsection{Attack on Reusing Counters for Non-Persistent Data} As described by Ye et al.\cite{OSIRIS}, assume an adversarial application uses known-plaintext and writes it to memory, however, if the memory location is non-persistent, the encrypted data will be written to memory but the counter might be not updated in memory yet. By observing the memory bus, the attacker can learn the encryption pad, i.e., OTP, by XOR'ing the observed ciphertext, $(E_{key}(IV_{new}) \oplus Plaintext )$, with the $Plaintext$. Even worse, it is also possible to predict the plaintext for initial accesses, for instance, zeroing at first access. To this end, the attacker knows the encryption pad, i.e.,  $E_{key}(IV_{new})$. Later, after recovering from a crash, the memory controller will fetch $IV_{old}$ and increment it, which generates  $IV_{new}$, and then uses it to encrypt the new data written to that location to become $E_{key}(IV_{new}) \oplus Plaintext2$. As the attacker knows the encryption pad, revealing the value of $Plaintext2$ only can be done by XORing the ciphertext with the previously observed encryption pad, i.e., $E_{key}(IV_{new})$. Note that the stale counter could have been incremented multiple times before the crash, hence multiple writes of the new application can reuse counters with known encryption pads. Note that such an attack only need a malicious application to run (or just predictable initial plaintext of an application) and having a physical attacker. %  if that memory location is assigned to a new application or use,

\subsection{Security Metadata Crash Consistency}

%Although crash consistency of security metadata has been rarely discussed in most memory encryption work\cite{SS,DEUCE}, it is extremely critical in persistent memory systems. 

\begin{figure}[htp!]
\begin{center}
   \includegraphics[scale=0.45]{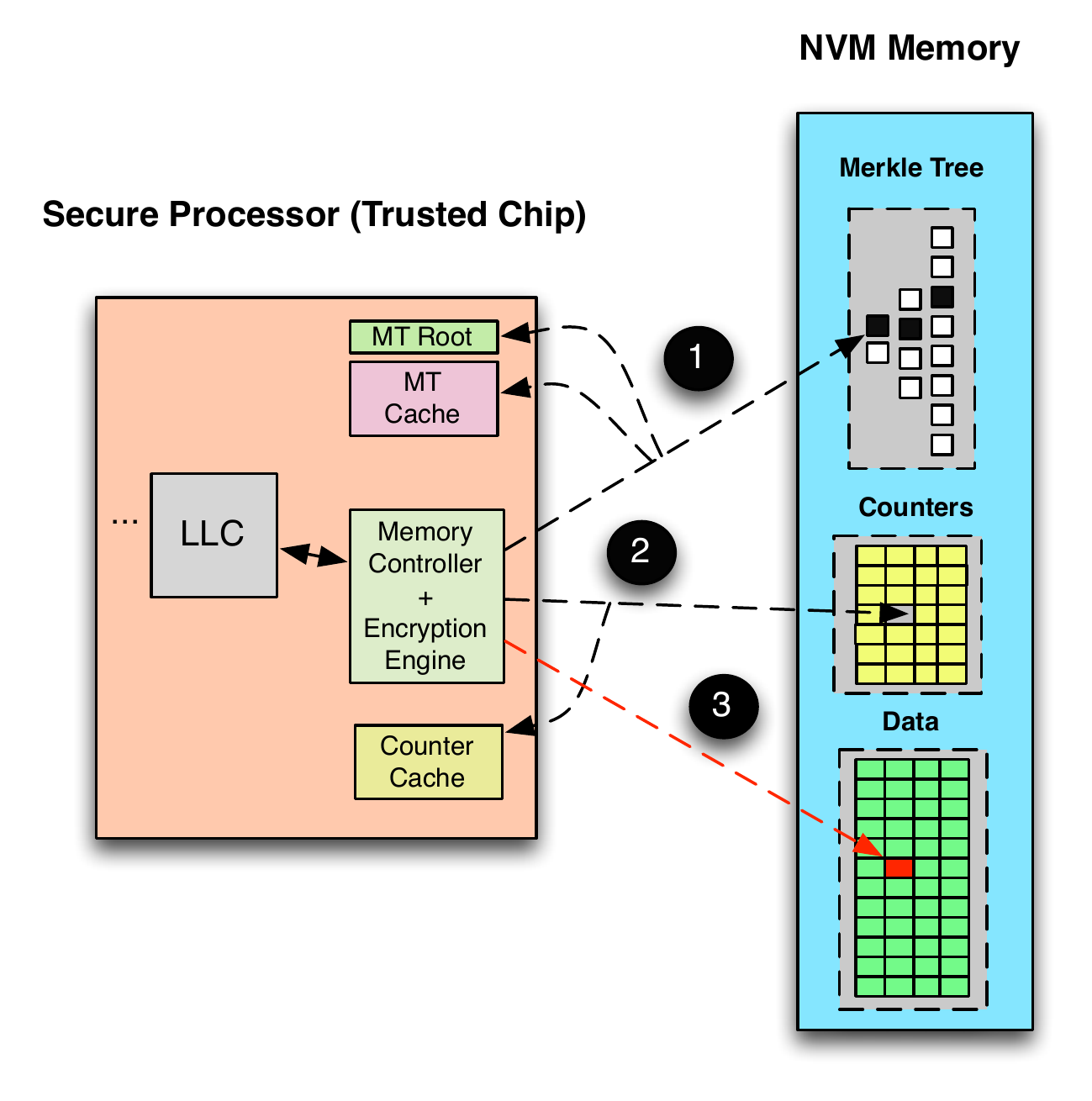}
  \caption{Description of write process that ensures crash consistency.}
  \vspace{-1em}
  \centering
  \label{fig:consis}
  \end{center}
\end{figure}

After a crash occurs, the system must be able to recover and restore its encrypted memory data along with its accompanying security metadata. Figure \ref{fig:consis} discusses the logical steps needed for crash consistency as described by Ye et al.\cite{OSIRIS}.

  %\vspace{-1em}

As depicted by Figure \ref{fig:consis}, the root of the Merkle tree (on-chip) should be updated/persisted (as shown in step \circled{1}) along with affected intermediate nodes inside the processor. However, only the \textit{root} of the Merkle Tree needs to be kept in the secure region. %Persisting intermediate nodes of Merkle Tree can speed up the integrity verification Persisting updates of intermediate nodes into memory is optional; it is feasible to reconstruct them from leafs (counters) and then calculate the root, will be verified through comparison with the one inside the processor.  %root of the Merkle Tree must persist safely across system failures, e.g., through internal processor NVM registers. Persisting updates to intermediate nodes of Merkle Tree after each access might speed up recovery time by reducing the time of rebuilding the whole Merkle Tree after crash, however, the overheads of such scheme and the infrequency of crashes make rebuilding Merkle Tree a more reasonable option.  %, we do not expect such crashes to occur often. will needlessly degrade performance, especially as we will need to rebuild and verify the whole Merkle Tree after recovering from crash. Most updates to intermediate nodes will only occur on cache due to high spatial locality of intermediate nodes; they cover large portion of memory. %Persisting changes to Merkle Tree intermediate nodes can reduce the recovery time significantly by relaxing the need to reconstruct the whole Merkle Tree, but only verification is needed after each access. 

In Step \circled{2}, the updated counter block is written back to memory as it gets updated in the counter cache. Counters are critical to keep and persist to avoid having the security of the counter-mode encryption be compromised. Note that even if the data is not expected to be recovered, the counters must be persisted to avoid repetition. While losing counter values can cause inability to restore encrypted memory data. Liu et al. \cite{HPCA_KHAN} observe that it is possible to only persist counters of persistent data structures (or subset of them) to enable consistent recovery. Unfortunately, this can lead to serious security vulnerabilities as discussed earlier; reusing counter values even for non-persistent memory locations can compromise the security of the counter-mode encryption. Moreover, modifying applications to expose their persistent ranges to the memory controller is challenging, especially for legacy applications. %Accordingly, a secure scheme that persists counter updates and does not require software alteration is in need. Note that, even for non-persistent applications, counters must be persisted on each update, otherwise the encryption key must be changed and all the memory must be re-encrypted with a new key. Moreover, if the persistent region in memory is huge, which is very likely in future NVM systems, most memory writes will naturally accompany with persisting the corresponding encryption counters. Hence step \circled{2} will be common.

Finally, in Step \circled{3}, the written data block will be sent to the memory. %As discussed earlier, some parts of Step \circled{1} and Step \circled{2} are critical for correct and secure recovery of encrypted NVMs. Updating the root of the Merkle Tree on the chip, updating the counter and writing the data are assumed to happen atomically, either through internal 3 NVM registers to save them before persisting them to memory or through small capacitor that is sufficient to complete 3 writes.
  %(starting filesystems started to  

\subsection{Motivation}
\begin{figure}[htbp!]
\begin{center}
  \vspace{-1em}
   \includegraphics[scale=0.3]{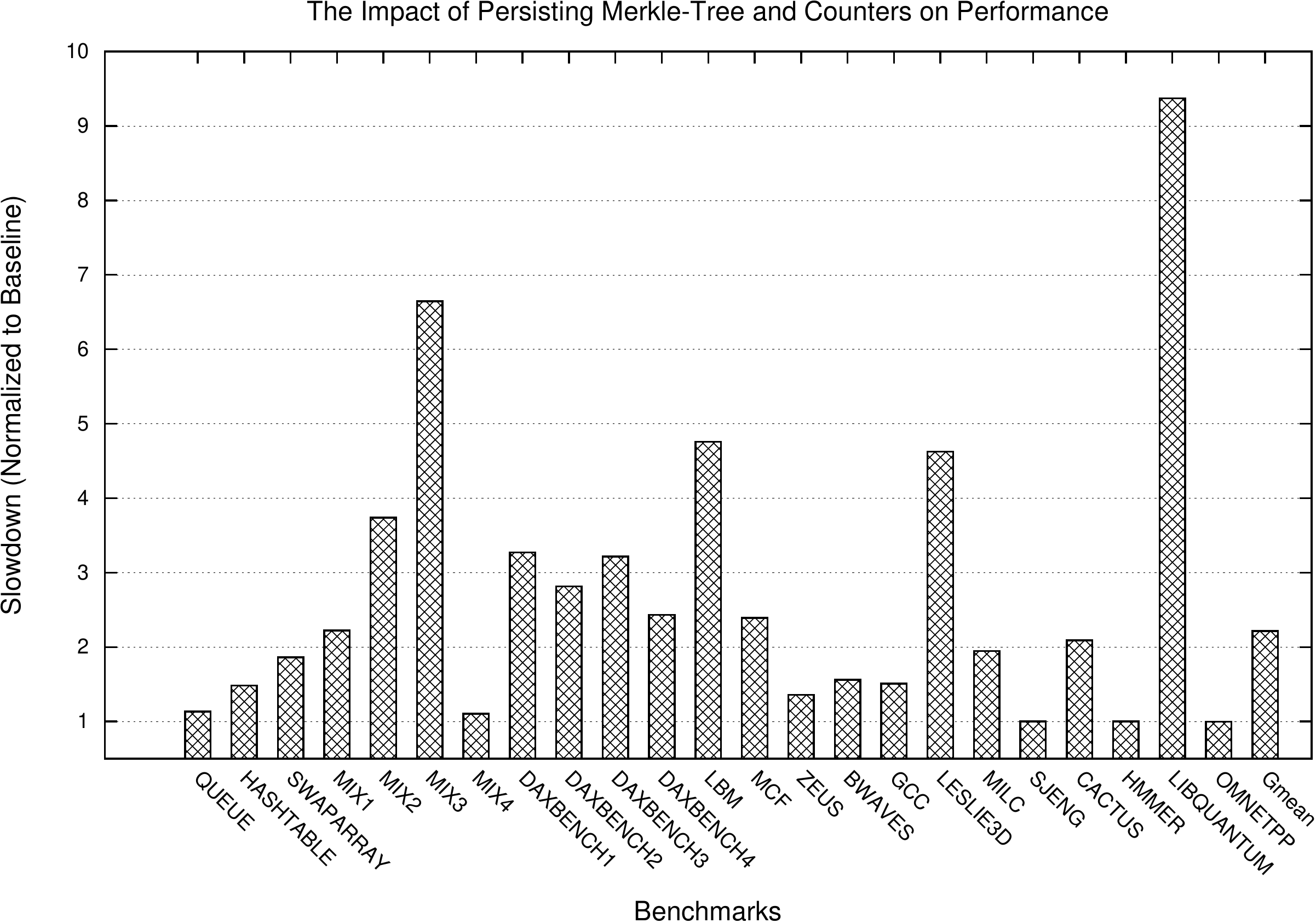}
  \caption{Performance Overhead of Persisting Security Metadata.} %State of the art counter mode encryption. AES is shown for encryption/decryption but other cryptographic algorithms are possible.}
  \vspace{-1em}
  \centering
  \label{fig:motiv}
  \end{center}
\end{figure}

As we now understand the issue of guaranteeing security metadata crash consistency, let's now discuss the impact of ensuring such consistency on performance of Non-Volatile Memories (NVMs). As mentioned earlier, a strictly persistent secure system would persist any changes to Merkle Tree and the corresponding encryption counter before each memory write operation. Figure \ref{fig:motiv}\footnote{More details about the workload and methodology can be found in Section \ref{sec:metho}} depicts the relative system throughput when using strict persistent security.  

As shown in Figure \ref{fig:motiv}, most workloads get severe performance degradation due to the impact of additional write operations to ensure persistency of security metadata. Notably, some workloads can be slowed down by as much as $9.4x$ compared to a system that does not guarantee such type of persistency. The impact of persisting security metadata depends mainly on the memory behavior and write-intensity of the running applications. Unfortunately, with such performance overheads, most users would avoid enabling persistent security which would leave their systems insecure and potentially their data unrecoverable. To address this issue, in this paper, we define the requirements of persistently-secure systems and study the impact of different schemes on performance, recovery time and resilience of the system. To the best of our knowledge, our paper is the first to discuss the issue of persisting both encryption counters and Merkle tree in NVM-based systems. 

%% file: design.tex
\section{Design}
\label{sec:design}

In this section, we first define the requirements of persistently secure systems. Later, we discuss the different design options that can be adopted for the purpose of implementing persistently secure systems. Finally, we discuss our proposed design that combines several novel optimizations.

\subsection{Persistently Secure Systems}

In order to start our discussion of the requirements of persistently secure system, let's first formally define such systems. 

\begin{theo}

A \textbf{Persistently Secure System} is any secure system that is capable of recovering its security metadata, e.g., encryption counters and Merkle Tree, in case of power-loss or system crash. Such a system should be able to verify that any (persistent) memory data being read after the crash reflects the most-recent value before or after system recovery. Specifically, if using counter-mode encryption, such a system should guarantee that one-time pads are never repeated regardless of the nature (persistent or non-persistent) of their corresponding data.

\end{theo}

As mentioned earlier, state-of-the-art designs for secure NVMs deploy counter-mode encryption along with an optimized Merkle Tree that protects tampering with data and encryption counters. Two key requirements for such schemes \circled{1} Never repeat the same counter without changing the key \circled{2} Ensure that the Merkle Tree root reflects the most-recent state of the memory. The hierarchical implementation of Merkle Tree is meant for reducing the overhead of verification in case of read operation; if you read a counter that has the most-recent value of its parent node being cached in the processor chip, i.e., has been already verified, then there is no need to go further (to upper levels) in the Merkle Tree. Similarly, if its grand-parent node is present but its immediate parent is not, then only the immediate parent needs to be fetched from memory and verified to match its hash value in its parent node and then the counter/leaf will be verified against its hash value in its fetched parent.

Verifying the counters being fetched from memory would require accessing all its parents up to the first one hits in the cache, which will be the root in the worst case. However, given the spatial locality of most applications, most applications rarely need to access more than few levels from memory. Unfortunately, the write operation is quite different; all upper levels must be updated including the root of the Merkle Tree. As NVMs are expected to have huge capacities, e.g., tens of TBs, the height of the Merkle Tree is expected to be in order of tens of levels, which would require updating tens of intermediate nodes blocks. Even worse, to achieve persistence of such security metadata, we need to persist each updated intermediate node block in the NVM. This brings us to the discussion of the requirements of implementing persistently secure systems.

\noindent \circled{1} \textbf{Encryption Counters:} While losing the most recent values of encryption counters of non-persistent data structures or data would likely not affect the correctness of the system, it can compromise its security as discussed earlier. Accordingly, to prevent such security vulnerability, encryption counters should be strictly persisted or there must be a guarantee of not repeating one-time pads. 

\noindent \circled{2} \textbf{Merkle Tree:} Persisting Merkle Tree updates is necessary to ensure the system ability of verifying encryption counters when recovered. However, we observe that not all parts of the Merkle Tree need to be persisted. Instead, just persisting the lowest levels of Merkle Tree is sufficient to rebuild the whole Merkle Tree at the time of recovery, however, at the cost of creating a single-point of failure and increasing recovery time as we will discuss later.

\subsubsection{Encryption Counters Persistence}

Encryption counters of persistent data must be updated strictly before updating the data to ensure consistent recovery of data. Meanwhile, as also observed by Liu et al.\cite{HPCA_KHAN}, counters of non-persistent data do not need to be strictly persisted. However, while prior work \cite{HPCA_KHAN} completely relaxes the persistence of such counters, we argue that we additionally need to ensure that such counters should never repeat with the same encryption key. Later, we will discuss a scheme that relaxes counters of non-persistent data while ensuring that they will never repeat. Meanwhile, the encryption counters of persistent data will be strictly persisted before updating their corresponding data.

\begin{obs}
Encryption counters of persistent data must be persisted and up-to-date during normal operation and across system failures/crashes. In contrast, for non-persistent data, the encryption counters must be up-to-date during normal operation and guaranteed not to repeat old values after recovering from crash.
\end{obs}

\subsubsection{Merkle Tree Persistence}

\begin{figure}[htp!]
\begin{center}
  \vspace{-0.4em}
   \includegraphics[scale=0.55]{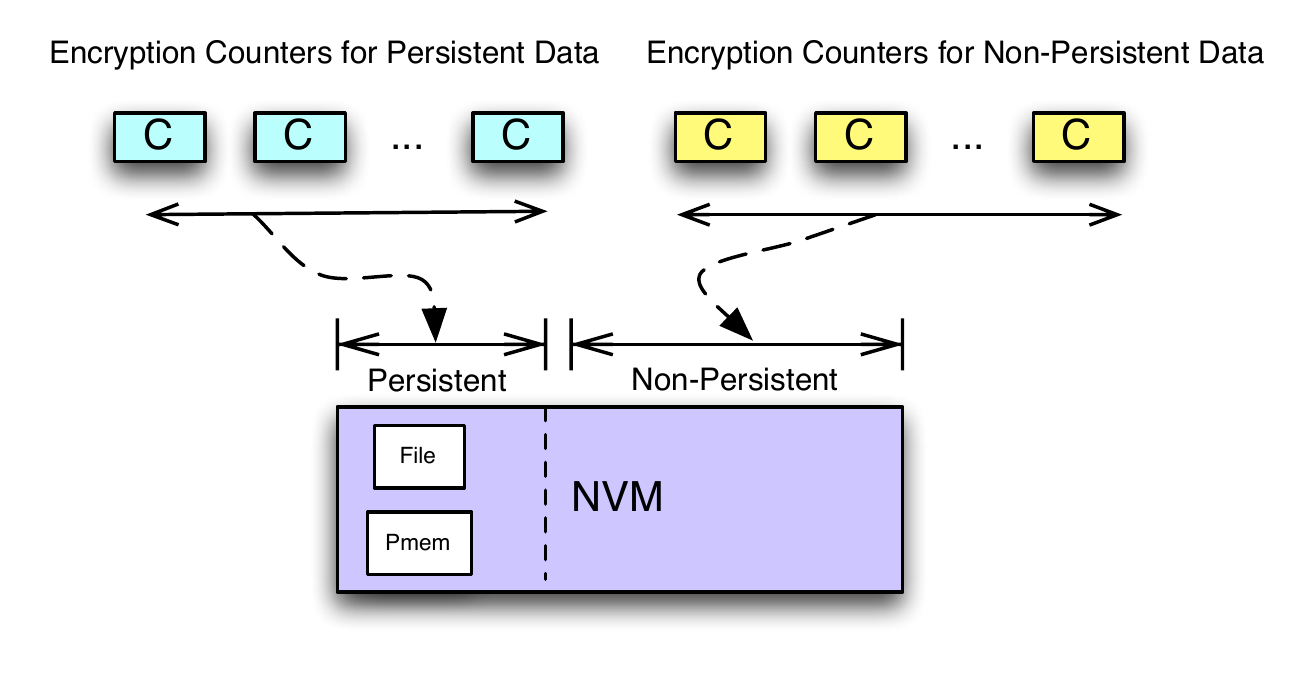}
  \caption{Encryption counters for data with different persistency requirements.}
  \centering
  \vspace{-1em}
  \label{fig:partition}
  \end{center}
\end{figure}

\begin{figure*}[htbp!]
\begin{center}
  \vspace{-0.4em}
   \includegraphics[scale=0.45]{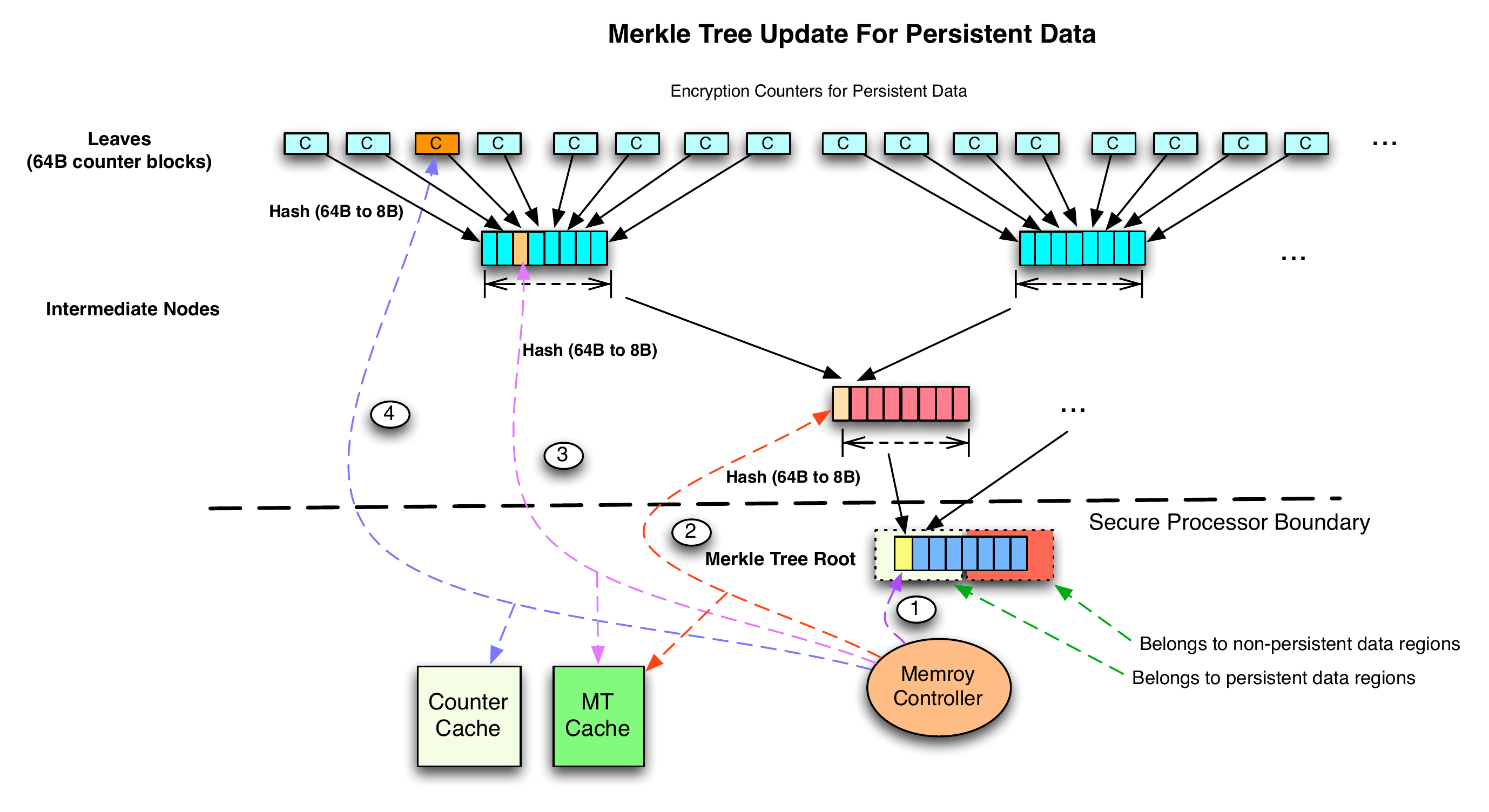}
  \caption{Updating Merkle for counters correspond to data with different persistency requirements.}
  \centering
  \vspace{-1em}
  \label{fig:MT_PERSIST}
  \end{center}
\end{figure*}

While Merkle Tree covers all encryption counters, regardless of their persistence requirements, not all parts of the Merkle Tree need to be persisted. Specifically, given that current systems define the persistent regions of the NVM during bootup time, there is a range in memory that is guaranteed not to be used for persistent data. For instance, if the Linux Kernel defines the first 4GB of memory to be used for persistent memory, i.e., can be mounted and used by PMEM and DAX applications, then other parts of the memory do not ensure any persistence of data. Thus, persisting the updates to Merkle Tree intermediate nodes that only cover non-persistent regions  might be relaxed as long as this ensures integrity during the normal operation. 

\begin{obs}
For non-persistent data, it is important to ensure their integrity during run-time, however, after a crash, we are no longer concerned about their values being lost or tampered with. In contrast, for persistent data, the data and counters integrity must be ensured during normal operation and across system failures/crashes.
\end{obs}

Based on the above observation, a Merkle Tree can be subdivided into multiple subtrees, i.e., some parts only cover the first 4GB of the main memory as aforementioned, hence only that part of the tree needs to be persisted strictly. In other words, the Merkle Tree can be vertically divided into persistent and non-persistent tree. We also observe that just persisting the leaves of the Merkle Tree is sufficient to allow its reconstruction, however, this can increase the recovery time and create single-point of failure, which would require additional fault isolation mechanisms.

\subsection{Relaxed Schemes}

In this part of the paper, we will discuss several optimizations and their potential impacts.

\subsubsection{Relaxed-but-Secure Persistence of Encryption Counters}

To understand this scheme, let's first discuss how current systems adopt persistent memory. In current systems, at the time of bootup, the kernel is initialized with a value that determines which range of the memory module is considered persistent. Such region can be formatted and mounted as a filesystem, e.g., ext4 filesystem. Additionally, such persistent region can be used by persistency APIs, such as PMDK, which would backup the persistent data structures by files in the persistent region. Any file in the persistent region can be memory-mmaped, i.e., {\tt mmap'ed}, and accessed through typical load/store operations. Meanwhile, the non-persistent region in the NVM device is not ensured recoverability after a crash; an application will only be able to access data persisted to the persistent region after recovery. Figure \ref{fig:partition} depicts a system with encryption counters of persistent and non-persistent regions in an NVM device. Note that such distinction between persistency of regions does not require modifying applications to explicitly define persistent data structures and exposure them to hardware as in prior work \cite{HPCA_KHAN}.

As we now understand how future NVM devices will be leveraged by OS, let's now discuss how encryption counters should be treated differently based on that. While encryption counters of persistent data must be recovered correctly for both correctness and security reasons, those of non-persistent data can be relaxed from ensuring correctness after recovery. The only requirement of encryption counters of non-persistent data is that they should not produce a previously used one-time pad, which we can ensure by using a different encryption key. %  used as long as the one-time-pad is never reused by using different keys.

Based on our observation that non-persistent data can lose their most-recent counter values after recovery, however, they should be never reused with the \textit{same key}, we devise a design that uses two different keys. One key, which we refer to by \textit{Persistent Key}, and the other key which we refer to by \textit{Volatile Key}. The persistent key is used for encrypting/decrypting persistent data regions, whereas the volatile key is used for non-persistent data. The volatile key will be changed after each system reboot or recovering from crash, which allows us to securely relax the persistence of encryption counters of non-persistent data. Our design only requires modifying the memory controller to be able to receive a command from the kernel which notifies the MC about the start and end address of the persistent region, and to use such information to decide which key to use. Moreover, the kernel needs to send the memory controller, through a memory-mapped register, a command to change the volatile key at the time of recovery or reboot.

\begin{obs}
In systems with persistent and non-persistent regions, if counter-mode encryption scheme is employed, separate keys can be used for each region. As encryption counters of the non-persistent region can repeat due to optional persistence necessity, the key used for such region must be changed after recovery or system reboot. By doing so, even if the same IV gets reused, the one-time pad will be different: $E_{key1}(IV) \neq E_{key2}(IV)$. Thus, updating encryption counters of non-persistent data can occur in counter-cache and use a write-back scheme instead of write-through.

\end{obs}

\subsubsection{Relaxed-Persistence Subtrees of Merkle Tree}

As we have discussed in the prior part, some parts of the memory are not expected to recover data correctly after a crash. However, such regions require that their integrity be protected during normal run-time, but do not have any expectation of integrity-protection after recovery from crash as the data will be discarded anyway. To better understand how an optimized scheme that relaxes persistence of non-persistent sub-trees of Merkle Tree, Figure \ref{fig:MT_PERSIST} shows how Merkle Tree can be updated based on the persistency requirement of the underlying data.

As depicted in Figure \ref{fig:MT_PERSIST}, when an update occurs for a persistent memory location, all the corresponding parts for such location in Merkle Tree at all levels should be updated up to the root (inside the processor chip). Updates to Merkle Tree for persistent data should be updating both the memory copy and the cache copy of \circled{1} The root of the Merkle Tree, \circled{2} The root's child node that owns the updated counter, and all other intermediate nodes correspond to the updated counter. Finally, as in \circled{4}, updating the counter in both the memory and counter cache. Note that, for non-persistent data, updating the intermediate nodes and counters in the cache and lazily write it back to memory is sufficient. Clearly, the subtrees that belong to persistent regions should be separable, i.e., some parts of the root only belongs to persistent-region and others only to non-persistent regions. Having some parts belong to both is challenging can lead to unverifiable parts after recovery; that part will reflect most-recent values of both types of data, but we will be unable of reproducing the root due to the lost updates of intermediate nodes and counters of non-persistent data. The only level that has MAC values on the same block for both persistent and non-persistent data is the root, however, they are clearly separable. To avoid cases of MAC values cover both types of data, the memory space ratio between persistent and persistent-data should 0:8, 1:7, 2:6, 3:5, 4:4, 5:3, 6:2, 7:1 or 8:0. In other words, each MAC value in the root should cover either persistent or non-persistent data but not both. Note that updates to the root does not need to be updated in memory as it is guaranteed to be persistent in the processor, i.e., in a NVM register inside the processor.

\begin{obs}
Merkle Tree can be logically divided into subtrees that either belong to persistent data or non-persistent data. Due to the architectural layout of the tree, the ratio of persistent to non-persistent data is limited to specific values that guarantee each MAC value in the root belongs to either persistent or non-persistent data. To this end, the Merkle Tree subtrees that belong to the non-persistent ranges can be updated in the MT cache and lazily updated in memory when written back.

\end{obs}

\subsubsection{Bottom-Up Merkle Tree Persistence}

As we discussed in the previous part on how to divide Merkle Tree vertically into persistent and non-persistent subtrees, we now will discuss which levels of Merkle Tree need to be updated and the impact of possible relaxation schemes.

Unlike encryption counters, Merkle Tree intermediate nodes can be rebuilt if lost and their main use is to speed up verification and updating the root of the Merkle Tree. However, if encryption counters are guaranteed to be strictly persistent and up-to-date, then after a crash it is possible to rebuild all levels of intermediate nodes all the way up to the root which needs to match the root value in the processor. While this looks like a straight-forward optimization, it can actually create a single-point of failure; if any counter has been corrupted, e.g., due to memory error, we can only know that the Merkle Tree root has not matched and thus none of the memory is integrity-verifiable. While a Merkle Tree root can hold 64B instead of only 8B values, still an uncorrectable counter error would result in $\frac{1}{8}$th of the memory being lost/unverifiable. As emerging NVMs are expected to contain terabytes of data, the chance that an uncorrectable error of any counter can lead to large part of memory being unverifiable is unacceptable. Moreover, only guaranteeing the persistence of encryptions would require high recovery time due to the need of iterating over all encryption counters to rebuild all levels of Merkle Tree. 

To enable high-resolution of identifying unverifiable locations in addition to keep the recovery time manageable, we propose persisting the first $N$ levels of the Merkle Tree (from bottom to up) while optionally maintaining several NVM registers within the processor. The value of $N$ depends on the acceptable performance/recovery-time trade-off. Persisting low levels of Merkle Tree can help isolating problems and identifying which counters are corrupted/uncorrectable. Moreover, more number of levels being guaranteed persistence allows shorter recovery time to reconstruct the Merkle Tree. Also note that since higher levels of the Merkle Tree has much less intermediate nodes, persisting them reduces the chances of the inability to construct Merkle Tree due to uncorrectable errors. 

\begin{obs}

Persisting parts of the Merkle Tree can avoid single-point of failures resulting from uncorrectable errors of encryption counters. Moreover, it reduces the Merkle Tree recovery time by reducing the number of levels need to be rebuilt after recovery. Internal NVM registers inside the processor can be also used to hold upper levels of Merkle Tree, e.g., root's immediate children and grand children (a total of 73 blocks), thus it can declare only a small percentage of memory unverifiable in case uncorrectable errors corrupt encryption counters or the persisted parts of the Merkle Tree.

\end{obs}

%Thus, we propose the following two schemes:

%\begin{itemize}

%\item \textbf{Hardened Leaves:} Each counter block will be at least duplicated and both versions will be persisted along with their ECC bits on each update. This mechanism eliminates the risk of having single-point of failure.

%\item \textbf{Second-level Persistence:} While the first approach can reduce the risk of single-point of failure, it can increase the Merkle Tree reconstruction time after recovery due the need to build the whole tree again. Meanwhile, if the immediate parents of the leaves are persisted as well, then the reconstruction time is reduced by $8x$ for a typical 8-ary Merkle Tree. 

%\end{itemize}

%Both of the above techniques can be combined together to ensure resilient and practical scheme for persisting only the low levels of Merkle Tree and relying on reconstructing the rest of the tree during recovery. Note that since reconstruction of Merkle Tree will start from second level, the construction starts from Level 3 (the grandparents of leaves) which has $8x$ smaller number of nodes and hence number MAC values to be produced. As the impact of recovery time is a function of how often does the system crash, we choose Level 2 to be the point of synchronization, however, upper levels can be also persisted in systems where shorter recovery time amortizes the performance hit. %gigabytes of memory, e.g., 128GB, immediate parent nodes of counters are only thousands of block, which makes Merkle Tree are sufficient to 

\subsection{Triad-NVM}
\label{sec:des}

Our design that includes all of the proposed optimizations is referred to as \textit{Triad-NVM}. Triad-NVM logically divides the Merkle Tree into subtrees, persistent and non-persistent. Moreover, it strictly persists counters that belong to persistent regions. Finally, for persistent regions, the Merkle Tree must be reconstructed after crash, thus it is necessary to ensure resiliency through persisting additional levels as configured by system owner or recommended by the system architects. For non-persistent regions, there is no need to ensure reconstruction of the corresponding subtree of the Merkle Tree, thus updates to such subtree are not strictly persisted.

At the recovery time, the Merkle Tree has to be reconstructed again to be able verify any tampering. During recovery, before admitting the current status of the system as verified, all persistent locations need to be verified through reconstructing the corresponding parts of the Merkle Tree and verifying that it generates a correct root. This ensures that the data has not been tampered with from (or before) crash time till recovery time. However, for non-persistent data locations, while we do not need to verify the integrity of data at the recovery time, we need to be able to verify it after recovery. Thus, Merkle Tree subtrees that correspond to non-persistent regions should be constructed. Now the question is how can we do that in an efficient way. 

\noindent \textbf{Reconstructing Subtree of Non-Persistent Region:} As mentioned earlier, in state-of-the-art schemes, e.g., Bonsai Merkle Tree \cite{MERKLEE}, the MAC values stored with data are calculated over data and encryption counters and hence all data and its accompanying MAC values need to be reinitialized based on the updated counter values after recovery. Note that since persisting updates to Merkle Tree and counters of non-persistent data is relaxed, there is no guarantee that the recovered counter values are similar to those used to calculate the MAC values accompanying their corresponding data. Accordingly, there will be inconsistency that will result in errors and mistakenly flagging tampering/errors during normal operations after recovery.

Unfortunately, it is very time-consuming to iterate over all non-persistent data to reinitialize counters, data and its MAC values. To better understand the overhead, let's assume a 6TB of NVM, where 50\% is used as non-persistent data, i.e., normal memory. At recovery time, if reading each data block and initializing it takes $100ns$, then just iterating over the 3TB (non-persistent region) would take 5154 seconds (almost 85.9 minutes). However, for the persistent data (the other 3TB), since the low-levels of Merkle Tree along with data and counters have been strictly persisted, only the upper levels of Merkle Tree. For instance, if only the first level of Merkle Tree (the parents of counter blocks), and reading each block and calculating its MAC values takes $100ns$, then that will take only 92 seconds (~1.5 minutes). For more aggressive persistence of Merkle Tree such as up to level 2 (the grandparents of counter blocks), then the construction time will take only 11.5 seconds.

\begin{obs}

Relaxing the persistence of Merkle Tree parts and counters that correspond to non-persistent data can result in significant increase in recovery time. Unfortunately, given the significant costs of system downtime (hundreds of thousands per minute \cite{COST}), completely reconstructing counters and Merkle Tree parts of non-persistent data at the recovery time can lead to unanticipated system unavailability.

\end{obs}

To avoid an unanticipated long recovery time of rebuilding counters and Merkle Tree parts of non-persistent regions, we additionally propose \textit{Lazy Recovery of Low-Levels of Non-Persistent Merkle-Tree}. As the number of counter and data blocks is the largest and they consume most of the rebuilding time, we can lazily update them as following. By initializing all intermediate nodes at level 1 (parents of counter blocks) with zeros, and constructing all upper levels sequentially, we can obtain an initial root value (or part of root) for non-persistent data. Later, any update for any counter value, if the parent intermediate node has a zero value, we do not flag an error or tampering as the upper levels of Merkle Tree (and root) are already updated to reflect such value, however, we know that this is the first write to the counter block after recovery and hence we zero out (or initialize) the counter value and update the parent node accordingly. Note that since each counter block has its MAC value stored in 8 bytes of the 64B parent node, only the corresponding 8B in the parent node would indicate if it is the first update after recovery or not. While the odds that a counter block value would naturally lead to 64-bit zero value is only $\frac{1}{2^{64}}$, to avoid falsely assuming an initialized counter value, if a counter block naturally has a MAC value of 0, we re-encrypt one of its cachelines and accordingly increment its minor value and calculate its new MAC value and use it to update the parent block if new MAC value does not equal to 0. By lazy update of non-persistent data and counter blocks, we ensure a recovery process that only needs to iterate over levels 1 or 2 instead of all corresponding data and counter blocks.

\begin{figure}[htbp!]
\begin{center}
  \vspace{-0.4em}
   \includegraphics[scale=0.2]{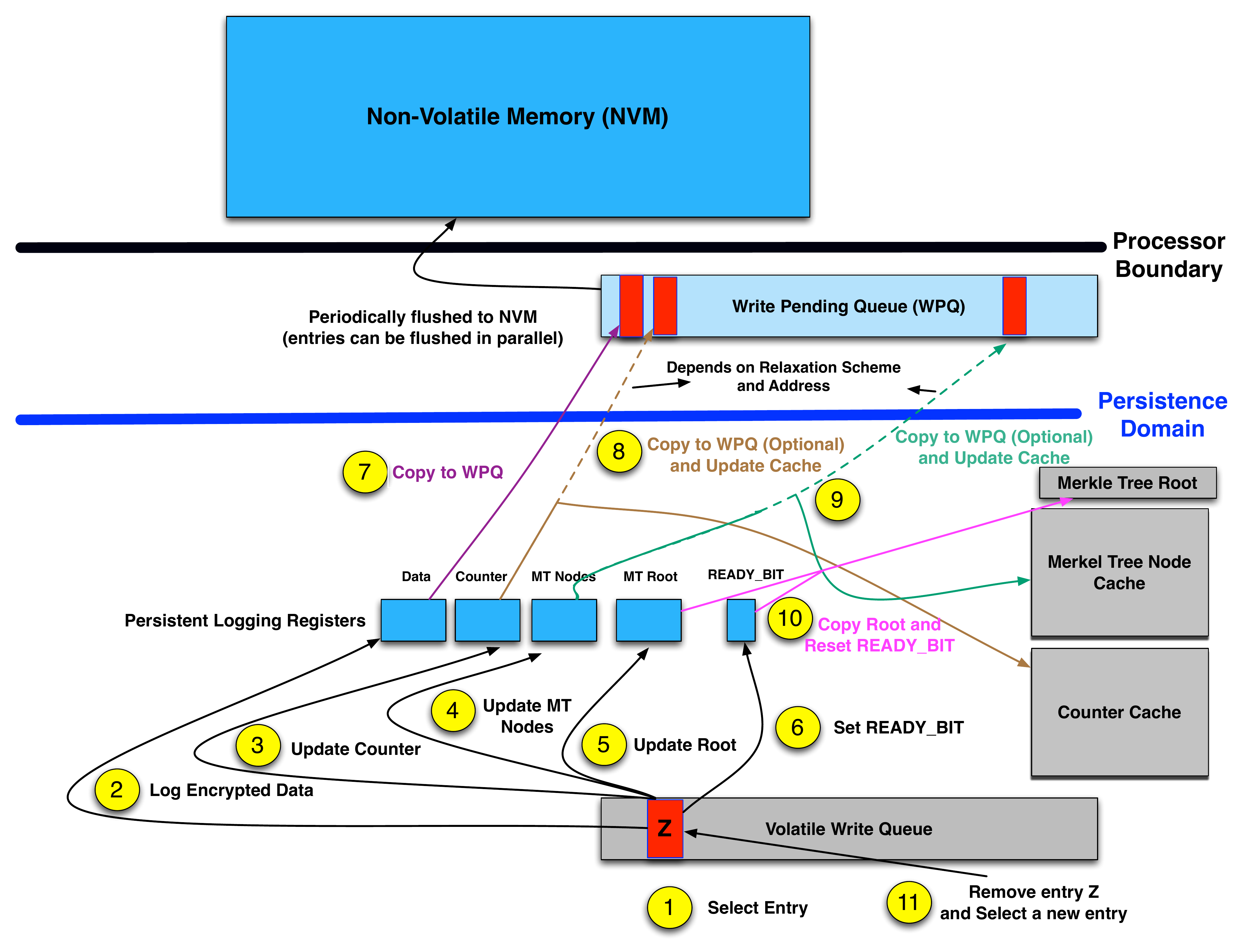}
  \caption{High-Level Overview of Write Operation on TriadNVM.}
  \centering
  \vspace{-1em}
  \label{fig:TRIADNVM}
  \end{center}
\end{figure}

Figure \ref{fig:TRIADNVM} depicts the Triad-NVM design. As mentioned earlier, Write-Pending Queue (WPQ) is considered part of the persistence domain (power-fail protected domain) in modern processors \cite{rudoff2016dpi}. Thus, anything reaches there should be consistent or there is a way to ensure its consistency. To do so, each write operation, before being persisted (removed from volatile buffer), it logs all its corresponding updates (counter, data, MT nodes and root) to persistent registers inside the processor and then set a persistent bit (called \textit{READY\_BIT}). If a crash occurs while copying the updates in persistent registers to WPQ, then when the system restores the memory controller will attempt to write the persistent registers to NVM (or WPQ) again. At the time of copying the contents from the registers to WPQ, Triad-NVM selectively chooses if the counter or Merkle Tree nodes should be copied to WPQ or it is just enough to update them in caches (as shown in steps \circled{8} and \circled{9}). Note that once a dirty block gets evicted from these caches it will go to WPQ as usual. Persistent registers can be implemented as fast NVM registers or volatile registers will be flushed to slower NVM registers once a crash occurs through leveraging ADR or residual power. The number of persistent registers depends on the Triad-NVM model, e.g., if TriadNVM-2 is used then we only need 5 registers, whereas if we use the impractical strict persistence then we might need up to 15 registers. Note that using persistent registers have been also assumed in state-of-the-art work \cite{OSIRIS}.

%Figure \ref{fig:TRIADNVM} depicts the high-level design of Triad-NVM. 
If the updated encryption counter corresponds to a persistent memory region, it will be strictly updated (copied to WPQ). However, for Merkle Tree, if the updated parts belong to a persistent memory region and in a level higher than the \textit{persist level}, then the updates will be persisted to NVM (copied to WPQ). Note that the persist level is the highest level of Merkle Tree that is guaranteed to be strictly persisted, e.g., if such a level is 2, then counters, their parents and grandparents are guaranteed to be persisted after each update. Note that TriadNVM recovery process is as simple as iterative over the intermediate notes at the persist level and construct the upper levels of the persistent memory regions, however, additionally initialize the intermediate nodes of such level to zeros for non-persistent memory regions before constructing their upper levels of the tree.

%% file: metho.tex
\section{Methodology}
\label{sec:metho}

In this section, we describe our evaluation methodology. Since our work involves kernel modifications and both persistent and non-persistent applications, we use Gem5 simulator\cite{GEM5} in its full-system mode. The kernel we simulate is based on Linux Kernel 4.14. Moreover, the disk image we use is based on Ubuntu 16.04 distribution. The kernel was initialized with configuring the 4GB starting from 12GB as a persistent region, e.g., the kernel was initialized with {\tt memmap=4G!12G}. Later, the persistent regions was formatted with DAX-enabled ext4 filesystem and mounted to be used by persistent applications and libraries. Table \ref{tb:config} presents the architectural configurations of our simulated system:

\begin{table}[htp!]
 \vspace{-1em}
\centering
\caption{Configuration of the simulated system.}
\label{tb:config}
\scriptsize
\begin{tabular}{|l|p{5.8cm}|} \hline
\multicolumn{2}{|c|} {\bf Processor} \\ \hline
CPU & 8-core, 1GHz, out-of-order x86-64 \\
\hline
L1 Cache &  private, 2 cycles, 32KB, 2-way, 64B block\\
\hline
L2 Cache &  private, 20 cycles, 512KB, 8-way, 64B block\\
\hline
L3 Cache & shared, 32 cycles, 8MB, 64-way, 64B block\\
\hline   
\multicolumn{2}{|c|} {\bf DDR-based PCM Main Memory} \\ \hline
Capacity & 16 GB\\ % device size is 1G, but it says there could be 32GB/channel, so I am not sure how this is measured.
\hline  
PCM Latencies & 60ns read, 150ns write \cite{PCM2} \\
\hline       
Organization & 2 ranks/channel, 8 banks/rank, 1KB row buffer, \newline
Open Adaptive page policy, RoRaBaChCo address mapping\\
\hline
DDR Timing & tRCD 55ns, tXAW 50ns, tBURST 5ns, tWR 150ns, tRFC 5ns \cite{PCM2,HPCA_KHAN}\\ %Only few parameters are mentioned in Lee's paper so the citation is not accurate. The tRAS is always overridden by the minimum of tRCD+tCL+tPR. Hence mentioning it as 35ns could be misleading. % Actually I used 72ns for tRAS but I avoid mentioning it here. The parameters we used here have been discussed with Dr. Awad.
& tCL 12.5ns, 64-bit bus width, 1200 MHz Clock\\
\hline 
\multicolumn{2}{|c|} {\bf Encryption Parameters} \\
\hline 
Counter Cache & 128KB, 8-way, 64B block\\
\hline 
Merkle-Tree Cache & 128KB, 8-way, 64B block\\
\hline
Merkle-Tree & 9 levels, 8-ary, 64B blocks on each level\\
\hline
\end{tabular}
\end{table}

To evaluate the impact of our proposed optimizations, we run 3 sets of workloads: persistent applications, non-persistent applications and a combination of both. For the benchmarks being used, following is the description of the major applications followed by Table \ref{tb:mix} which shows the mixed workloads we use. For non-persistent workloads, we use representative benchmarks from SPEC2006\cite{SPEC}. We also implement persistent Hashtable, ArraySwap and Queue benchmarks using Intel's PMDK library. Additionally, we implement synthetic benchmark, DAXBENCH-$S$-$RW$ that leverages DAX to {\tt mmap} a file directly in the persistent region and access it through memory load/store operations with $S$ stride and $RW$ read to write ratio. Finally, we use those benchmarks to create multi-programmed workloads to enable studying the optimizations that relax persistence of security metadata for non-persistent data applications. 

\begin{table}[htp!] 
 \vspace{-1em}
\centering
\caption{Mixed Benchmarks}
\label{tb:mix}
\scriptsize
\begin{tabular}{|l|p{5cm}|} \hline
\multicolumn{2}{|c|} {\bf Benchmark Combination} \\ 

\hline 
DAXBENCH1 &  DAXBENCH-128-2 \\

\hline 
DAXBENCH2 &  DAXBENCH-1024-2 \\

\hline 
DAXBENCH3 &  DAXBENCH-256-2 \\

\hline 
DAXBENCH4 &  DAXBENCH-512-3 \\

\hline 
MIX1 &  Array-swap, Queue, Hashtable, DAXBENCH-64-2 \\
\hline 
MIX2 &  Mcf, Queue, Hashtable, DAXBENCH-64-2 \\
\hline 
MIX3 &  Mcf, LBM, Hashtable, DAXBENCH-512-2 \\
\hline 
MIX4 &  Array-swap, Hashtable, Hashtable, DAXBENCH-1024-2 \\
\hline 
% 9 &  DAXBENCH \\
% \hline
% 10 & DAXBENCH \\
% \hline 
% 11 & Array-swap, Hashtable, Mcf, LBM \\
% \hline 
% 12 & Queue, Hashtable, LBM, Libquantum \\
% \hline 
% 13 & 1 Array-swap \\
% \hline 
% 14 & 1 Queue \\
% \hline 
% 15 & 1 Hashtable \\
\end{tabular}
\end{table}

All applications were fast-forwarded to representative regions and each run simulated at least 200M instructions.

%% file: evaluation.tex
\section{Evaluation}
\label{sec:eval}

In this section, we evaluate our proposed optimizations and study the impact of combining them on Triad-NVM. We will first study the impact of each optimization on performance, and later study and analyze the trade-off between recovery-time and performance overhead when relaxing the persistence of Merkle Tree.

\subsection{The Impact of Relaxation Schemes on Performance}

\begin{figure}[htp!]
\begin{center}
  \vspace{-0.4em}
   \includegraphics[scale=0.35]{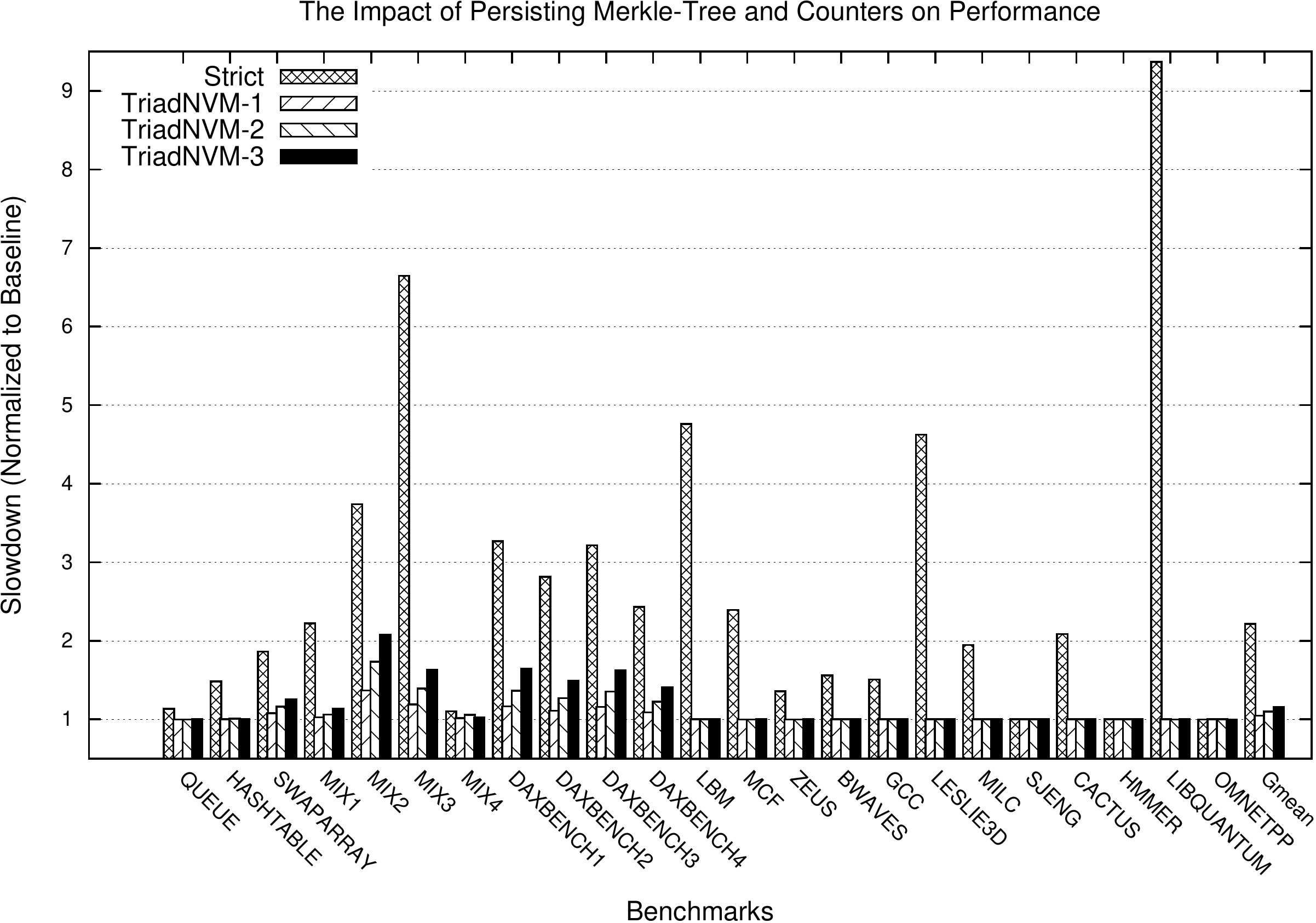}
  \caption{The impact of Merkle Tree persistence model on Throughput.}
  \centering
  \vspace{-1em}
  \label{fig:eval1}
  \end{center}
\end{figure}

Figure \ref{fig:eval1} shows the impact of Triad-NVM on performance. TriadNVM-$N$ represents TriadNVM with strict persistence of persistent regions up to the $N$th level of the Merkle Tree. As expected, for the persistent region, TriadNVM-1 is expected to perform better than TriadNVM-2 and TriadNVM-3. The main reason for that is that less number of writes need to occur to memory to ensure strict persistence of Merkle Tree, i.e., in TriadNVM-1 only the parent of the updated counter block needs to be persisted along with the counter block, whereas in TriadNVM-2 both the parent and the grandparent of the updated counter block need to be persisted along with the counter block. Clearly, there will be no difference between different TriadNVM persistence levels for non-persistent workloads, e.g., LBM and MCF. Moreover, in mixed workloads, only the writes to persistent regions will make difference between TriadNVM models.

Our results show that using strict persistence scheme can lead to an average of 2.21x slowdown, whereas TriadNVM-1, TriadNVM-2 and TriadNVM-3 lead to only 4.9\%, 10.1\% and 15.6\% performance overheads, respectively. Moreover, we observe that write-intensive workloads with non-persistent region allocations, e.g, Libquantum, can get an almost order of magnitude speed up when using schemes that are aware of persistent regions and thus relax the requirements of non-persistent memory ranges.

As mentioned earlier, one major shortcomings of the emerging NVMs are their slow writes and limited write endurance. However, while ensuring higher level of persistence would possibly reduce the recovery time and also avoid single-point of failures, it can excessively increase the number of writes. However, more relaxed schemes, e.g., TriadNVM-1, would lead to additional recovery time and potentially single-point of failure, however, it incurs smaller number of extra writes.

\begin{figure}[htp!]
\begin{center}
  \vspace{-0.4em}
   \includegraphics[scale=0.35]{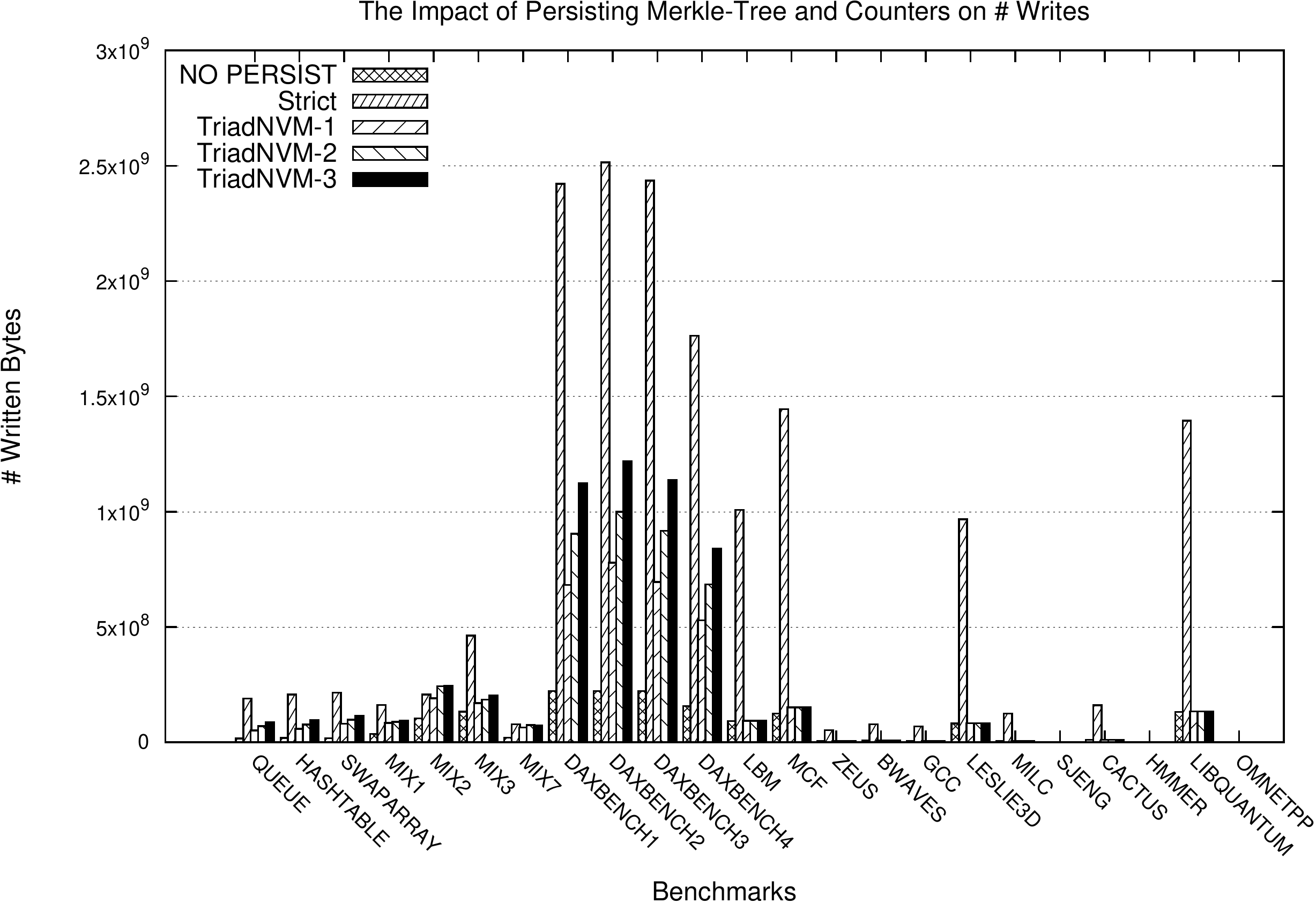}
  \caption{The impact of Merkle Tree persistence model on \# Writes.}
  \centering
  \vspace{-1em}
  \label{fig:eval2}
  \end{center}
\end{figure}

Figure \ref{fig:eval2}, shows the total number of times in our simulations for different schemes for each workload. We can observe that most applications' \# writes are proportional to the persistence level, however, since large percentage of writes could be resulting from natural cache write backs and data persistence, only the writes result from persisting security metadata would increase linearly with the level of persistence. We can observe that for most workloads, the number of writes for TriadNVM is close to the original number of writes without persisting security metadata.

\subsection{The Impact on Recovery Time}

\begin{figure}[htp!]
\begin{center}
  \vspace{-0.4em}
   \includegraphics[scale=0.34]{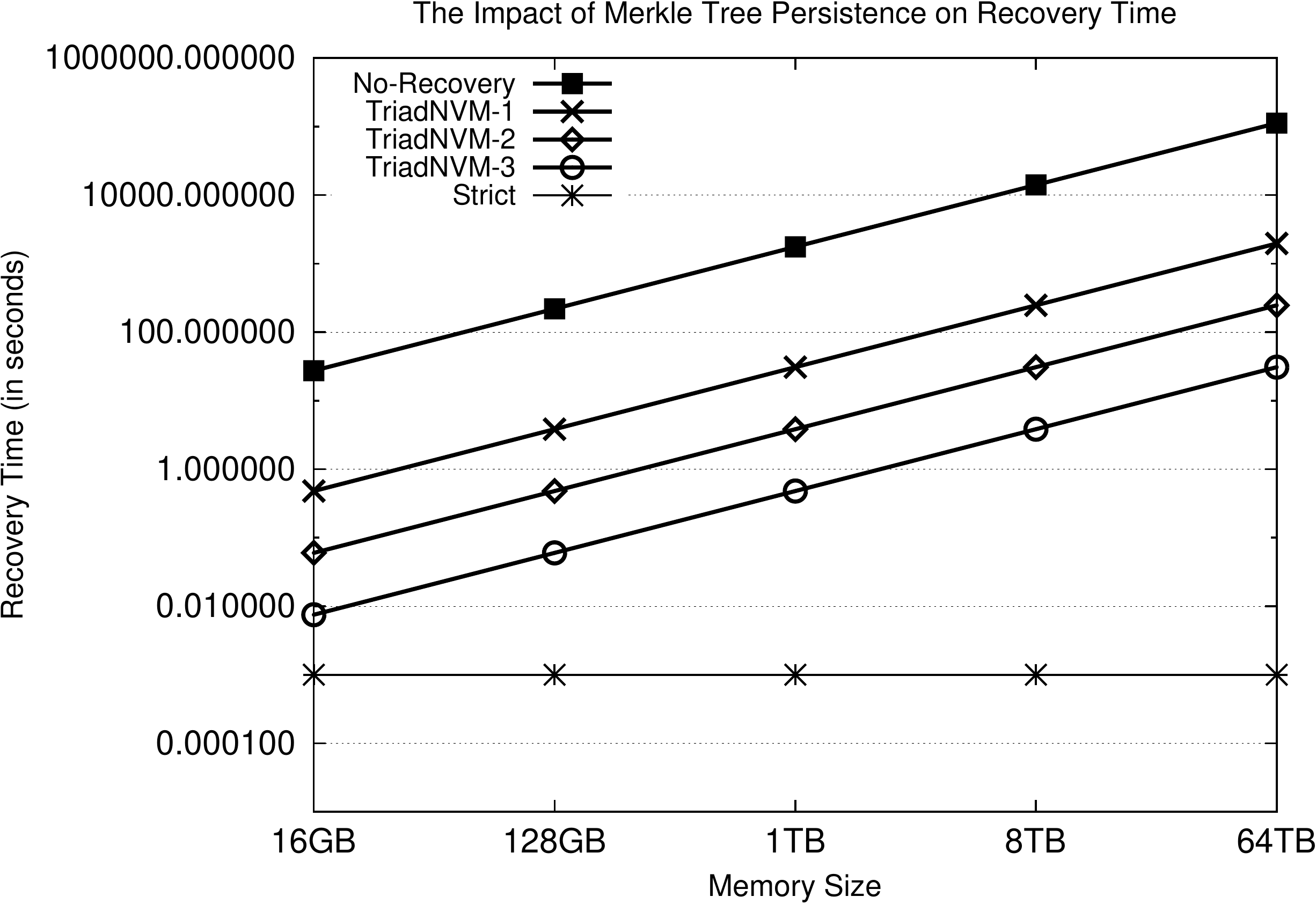}
  \caption{The impact of Merkle Tree persistence model on Recovery Time.}
  \centering
  \vspace{-1em}
  \label{fig:recovery}
  \end{center}
\end{figure}

As mentioned earlier, strict persistence ensures near zero recovery time, however, at the cost of significant performance degradation during normal operation. Meanwhile, not persisting security metadata at all will require creating Merkle Tree and initializing MAC values in data blocks which requires iterating over all memory blocks. In contrast, Triad-NVM guarantees the persistence of the counters and the first level of the Merkle Tree.  To estimate recovery time, we assume that reading a tree block in addition to calculating its MAC value takes $100ns$. Figure \ref{fig:recovery} shows the impact of persisting security metadata on recovery time.

As emerging NVMs are expected to be in terabytes, our goal is to get an estimation of how long does it take to recover the whole Merkle Tree for different persistency models. We expect TriadNVM-3 to have the least recovery time as it requires the least number of computations; starting the reconstruction from Level 3. Meanwhile, TriadNVM-2 brings in a tradeoff between better performance but almost ~8x slowdown. In the case of strict persistency, there is almost zero recovery time; all parts of Merkle Tree are up to date. As shown in the figure, the problem becomes more obvious when we deal with terabytes of memory, e.g., at 1TB the recovery (construction) time for no-persist scheme is 30 minutes. In contrast, for TriadNVM-1 it takes only 30.68 seconds to complete recovery. As system downtime can be in hundreds of thousands per minute \cite{COST}, some high-availability systems might trade performance for faster recovery time. For instance, if TriadNVM-2 or TriadNVM-3 are used with 1TB of memory, then the recovery time is 3.83 and 0.48 seconds, respectively. We can also observe that when the memory 8TB, the recovery time is almost 8x slower than that of 1TB, due to linear increase in number of nodes/blocks to iterate through to calculate MAC values. While TriadNVM can be configured to persist up to any specific value, we leave the decision of which level to persist to up to system integrator or administrator and based on the acceptable performance/recovery-time trade-off. Note that such a decision can be also affected by the crash rate and possibility of system failures in addition to how sensitive the data is and tolerance to declaring them unverifiable at recovery time. 

As resilience of the system is also important, TriadNVM-2 and TriadNVM-3 provide a higher resolution of pinpointing errors. In TriadNVM-2, if calculating the MAC values starting from level 2 does not eventually generate the root value stored in the processor, then TriadNVM restarts form level 1 and find out which node in Level 2 does not match with that calculated from its children, and hence only declaring the corresponding 32KB as unverifiable. Note that the root must match with that generated from those in Level 1, before declaring that Level-2 node to be the source of error. Similarily, if an uncorrectable error occurs in Level-1 nodes, while Level-2 nodes ended up generating the same root value, i.e, verified, then we can use Level-2 nodes to pinpoint which Level-1 node is the corrupted one. TriadNVM-3 follows the same logic but now add an additional level of isolation in case uncorrectable errors occurred on both Level 1 and Level 2. It is also important to note that since the number of levels of Level-2 and Level-3 are 8 and 64 times, respectively, smaller than that of Level-1, then the chances of uncorrectable errors on them is also smaller.

%% file: relat.tex
\section{Related Work}
\label{sec:relat}

Persisting security metadata has been rarely studied up until recently. Recent work \cite{HPCA_KHAN} proposed selective persistence of encryption counters without any discussion of how Merkle Tree persistence can be handled. Moreover, as discussed earlier, selective encryption counter is vulnerable to known-plaintext attacks as it allows the reuse of encryption counters for non-persistent data. In contrast, our work takes a more holistic approach of studying all relevant security metadata and discuss their persistence impact on recovery-time, resilience and performance. Additionally, we propose solutions to mitigate such overheads and to protect against potential repetition of one-time pad. 

Recently, Zuo et al.\cite{SECPM} proposed combining multiple updates of encryption counter block into one write to memory, especially for large transactions. Unfortunately, such a solution works only well if we can predict the spatial locality of writes to memory, which is typically hard to predict as they can result and interfere with evictions/write-backs from the Last-Level Cache (LLC). Thus, such a solution would be beneficial only when many writes to memory are spatially contagious and occur at relatively close time. In contrast, our solutions are generic and do not expect any special behavior from workloads. Additionally, we investigate Merkle Tree persistence, which has not been discussed or explored in any of the previous work we are aware of. Nevertheless, SecPM can be orthogonally augmented with our proposed solutions. Another recent work explore repurposing Error-Correcting Codes (ECC) to be additionally used as a sanity-check for the encryption counter \cite{OSIRIS}. By doing so, Ye et al.\cite{OSIRIS} propose a novel scheme that can be used to relax the persistence of encryption counters and rely on restoring it through trying several values until ECC match or indicate natural number of errors. Our work is orthogonal to Osiris and can be integrated with it to reduce the number of writes to persist counters in the persistent memory regions. 

Persisting data in NVMs have different models based on the required software and hardware changes\cite{NVHEAP,THYNVM,LLP,KILN}. Several APIs and libraries have been developed for the purpose of persisting data in emerging NVMs and utilizing the OS support for PMEM, e.g., Intel's PMDK \cite{PMDK}. In this paper, our support focuses on persisting security metadata accompanying persistent data when being written to NVM, and thus we do not require any changes at application or library level, hence Triad-NVM can be integrated with most persistency models.

%\noindent \textbf{Memory Persistency:} As a new technology possessing the properties of main memory and storage, persistent memory has the promising recovery capability and the potential to replace main memory. 
%Exploting the persistency feature of NVMs has realized through both hardware and software research efforts\cite{NVHEAP,THYNVM,LLP,KILN}. Pelley \cite{MP} refers, memory persistency, to constraints of write order in respects to failure and proposes several persistency models in either strict or relaxed way. DPO \cite{DPO} and WHISPER \cite{WHISPER} propose different persistent framework using strict and relaxed order respectively with different granularity. In our design, we assume conventional persistence model, i.e., cacheline flushing ordered by store-fencing, e.g., {\tt CLWB} then {\tt SFENCE}, to persist memory locations \cite{INTELSDM}.

Other works that target reducing NVM writes have not considered the problem of persisting security metadata \cite{SS,DEUCE}. All of these solutions can employ TriadNVM to ensure persistence and recoverability of security metadata.

%% file: concl.tex
\section{Conclusion}
\label{sec:concl}

In this paper, we propose TriadNVM, a set of schemes that enable efficient, resilient and performance-friendly recovery mechanism for security metadata. TriadNVM ensures the persistence of Merkle Tree and encryption metadata while incurring minimial overheads. Moreover, TriadNVM can be customized to different modes based on the desired recovery-time/performance trade-offs.